\documentclass[aip,twocolumn]{revtex4}
\usepackage{graphicx,color}

\begin{document}

\title{Extraction-Controlled Quantum Cascade Lasers}
\author{Andreas Wacker}

\email[]{Andreas.Wacker@fysik.lu.se}
\affiliation{Mathematical Physics,
Lund University, Box 118, 221 00 Lund, Sweden}

\date{July 29, 2010}
\begin{abstract}
A simple two-well design for terahertz quantum cascade lasers is proposed
which is based on scattering injection and the
efficient extraction of electrons from the lower laser level by resonant
tunneling. In contrast to existing designs this extraction also controls
the positive differential conductivity.
 The device is analyzed by calculations based on nonequilibrium Green
functions, which predict lasing operation well above 200 K at a frequency of
2.8 THz.
\end{abstract}



\maketitle


Since their first realization\cite{FaistScience1994}, 
quantum cascade lasers (QCLs)
have turned into versatile devices. While the first lasers operated 
in the infrared region above the optical phonon frequency 
(around 9 THz in the most common III/V semiconductor materials used), 
THz-QCLs, operating below this frequency, could be achieved later.
\cite{KohlerNature2002}. However the operation of these THz-QCLs has only been
established for temperatures up to 186 K yet\cite{KumarAPL2009a}. The
achievement of higher operation temperatures is of high significance 
for many technical applications,  as one could apply simpler cooling 
techniques.

Based on a single concept QCLs operate over a range of almost two orders of
magnitude in frequency from 1.2 THz (250 $\mu$m) 
\cite{WaltherAPL2007} to 114 THz (2.63 $\mu$m)
\cite{CathabardAPL2010}
except for a gap around the optical phonon frequency.
The basic ideas, some of them essentially  
going back to Kazarinov and Suris\cite{KazarinovSPS1971} are:
(i) The use of {\em electronic subbands} in semiconductor
  heterostructures as {\em upper} (subscript $u$) and {\em lower} (subscript $l$)  laser level
  allowing for the wide variation of the transition energy by properly chosen
  heterostructures.
(ii) Electric pumping by a bias along the growth direction of the
  heterostructure. Here the flow of current through the structure feeds
  electrons into the upper laser level coming from the 
  {\em injector level} (subscript $i$). While the further propagation from the
  upper laser level at 
  energy  $E_u$ is essentially blocked by a gap in the energy spectrum
  of the heterostructure, efficient pathways are provided for the
  emptying of the lower level into an {\em extraction level} (subscript $e$).
(iii) While  a single intersubband transition is typically not sufficient to
overcome the losses in the waveguide,  a {\em cascade} of identical structures
(called period) is realized, all contributing to the gain of the optical 
mode in the waveguide.

\begin{figure}
\includegraphics[width=8.3cm]{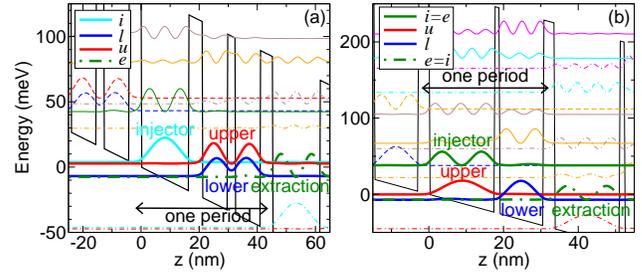}
\caption{(Color online) (a) absolute square of the 
Wannier states of the QCL from
 Ref.~\cite{KumarAPL2009a} together with the heterostructure potential.
(b) Same data for the device proposed here. The layer sequence is
$17.5/\underline{\bf 1.5}/11.5/{\bf 3}$ nm with effective masses of 0.067 in the
wells and 0.0919 in the barriers (bold symbols). The conduction band offset is
0.27 eV, which relates to a GaAs/Al$_{0.3}$Ga$_{0.7}$As structure. The
underlined barrier is doped with $3\times 10^{10}/\textrm{cm}^2$.
(Note that Wannier states  are not the commonly plotted
eigenstates of the Hamiltonian but are better localized within the period \cite{LeePRB2002}.)}
\label{FigWannier}
\end{figure}

As an example the Wannier states for a recent THz QCL with high-temperature
operation \cite{KumarAPL2009a} have been displayed in Fig.~\ref{FigWannier}(a).
Here the current flows via resonant tunneling from the injector into the
upper laser level and is extracted from the lower laser level by a further
resonant tunneling processes, while this extraction level is emptied by
optical phonon scattering. This {\em resonant phonon extraction} design
\cite{WilliamsAPL2003}, 
has been proven to be very effective for THz-QCLs as it selectively empties
the lower laser level while the upper laser level is less affected 
due to the resonance condition. 

An important issue for the operation of quantum cascade lasers is the
achievement of positive differential conductivity (PDC) 
for each period. Otherwise, for negative differential conductivity (NDC),
the electric field distribution becomes inhomogeneous over the structure
\cite{LUPRB2006}
and prevents from achieving the designed resonances in each period of 
the cascade. This
phenomenon of domain formation is well known from the study of superlattices (see \cite{WackerPR2002}
and references cited therein) where it constitutes the main obstacle to
observe the gain properties. 
Tunnel resonances exhibit commonly a current peak when the
levels align. This provides PDC/NDC if
the left level is located below/above the right level, respectively
\cite{CapassoAPL1986,KristensenSST1998}.
In particular this relates to the gain transition in THz
structures, where the energy separation is only slightly larger than the 
broadening and
thus constitutes a source of NDC. This
has to be compensated by PDC contributions in different parts of the
current flow through each period which is typically achieved by 
a second tunneling transition just below resonance. 
In most cases this is the transition
between the injector and the upper laser level, see also 
Fig.~\ref{FigWannier}(a). Albeit establishing record temperature operation
such a {\em tunneling injection} design has two shortcomings. (i) As the tunneling resonance should be
an effective source of PDC the injector level must 
exhibit an occupation at least comparable to 
the upper laser level. This restricts the possible inversion for a given total
carrier density. (ii)
Tunneling from the injector to the lower laser level constitutes a second
resonance which is a further source of NDC and thus of particular concern for
low lasing frequencies, when it mixes with the tunneling resonance into the
upper laser level. A possible solution of this problem is the development of
{\em scattering injection} designs
\cite{YamanishiOE2008,YasudaAPL2009}. Such a structure
was recently shown to exhibit improved temperature performance in the THz region\cite{KumarCLEO2010A}.
In these structures a tunneling resonance is included in the current flow
before the electrons reach the injector state in order to guarantee PDC.

Here a new structure based on scattering injection
is proposed where this tunneling resonance is skipped,
see Fig.~\ref{FigWannier}(b).
Instead, the tunneling resonance from the lower laser level to the extraction
level controls the current. The idea is that for biases below the designed 
operation point the carriers are essentially located in the lower laser
level. At the design bias the extraction level removes these carriers
effectively. Simultaneously, this level serves as the injection level for
the upper laser level of the next period via phonon scattering. Thus this
structure is a simplified combination between the resonant phonon extraction
and the scattering injection scheme, where both features are provided by the
same levels. This allows a design with only two wells per period
and thereby increases
the number of possible periods in the waveguide. (Two-well designs have been
already established for the conventional tunneling injection design
\cite{KumarAPL2009b,ScalariOE2010}.)

The design of the structure was optimized by 
calculations within the
nonequilibrium Green function model described in 
\cite{LeePRB2002,BanitAPL2005,LeePRB2006}
using an improved treatment of acoustic phonon scattering and including alloy
scattering, see \cite{NelanderDiss2009} for details. This model allows for a
consistent treatment of coherent evolution and scattering including level
broadening, and has been recently used by other groups as well \cite{KubisPRB2009,SchmielauAPL2009}.
Here the following issues were found to be of relevance for the final design:
(i) At the operating bias, the extraction level is in resonance with the lower
laser level and  located about one optical phonon energy above the upper laser
level of the subsequent period.
(ii) The higher levels do not provide further level spacings comparable to the
lasing transitions in order to avoid reabsorption at higher temperatures, when
they are partially filled.
(iii) Increasing doping enhances the number of carriers in the gain transition
but also strengthens impurity scattering associated with a larger linewidth and
a shorter lifetime of the upper laser state. The chosen doping was found to
provide the strongest gain.
(iv) Compensation effects \cite{BanitAPL2005} reduce the width of the gain
spectrum if the same doping atoms affect both laser levels. Thus the placing
of the doping in the barrier between the lasing states is advantageous.

\begin{figure}
\includegraphics[width=8.5cm]{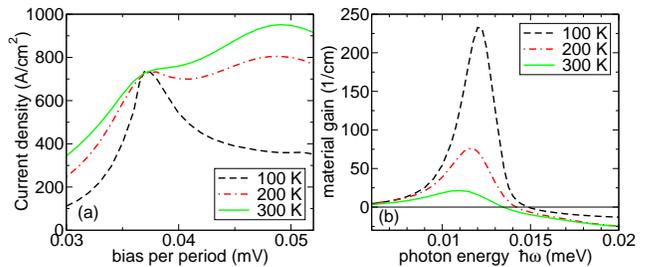}
\caption{(Color online) Simulation results for the proposed
  structure of Fig.~\ref{FigWannier}(b) at different lattice temperatures.
(a) Current versus bias per period. (b)  Gain spectrum at a bias of 
48 mV per period}
\label{FigKennGain}
\end{figure}
 
Fig.~\ref{FigKennGain}(a) shows the calculated current-voltage characteristics
for different temperatures for the optimized structure.
The currents are of the same magnitude as in the 
design of \cite{KumarAPL2009a} and thus the same thermal management should
work. The gain spectrum, see Fig.~\ref{FigKennGain}(b), shows a peak around
12 meV (2.8 THz). At 200 K the gain maximum is 76/cm which is almost twice
as large as the calculated value for the  structure of \cite{KumarAPL2009a}
(42/cm at 17 meV for 200 K). Given the fact that the latter sample exhibited
laser operation until 186 K, laser operation well above 200 K can be
expected for the new design. At 300 K the peak gain is reduced to 22/cm, which
is most likely not sufficient to overcome the waveguide losses.

\begin{table}
\begin{tabular}{|l|c|c|c|}
\hline
temperature (K) & 100 & 200 & 300 \\ \hline
upper $n_u$ ($10^{9}/\mathrm{cm}^2$) & 22.3 & 16.6 & 13.5 \\
lower $n_l$ ($10^{9}/\mathrm{cm}^2$) & 5.5 & 8.7 & 9.9 \\
$n_ue^{-\hbar\omega_{\mathrm opt}/k_BT}$ ($10^{9}/\mathrm{cm}^2$) & 0.3 & 2.1 & 3.4 \\
FWHM of gain spectrum (meV) & 2.2 & 2.9 & 3.9 \\
Peak gain (1/cm) & 233 & 76 & 21 \\ \hline
\end{tabular}
\caption{Calculated quantities for the gain transition at 48 mV per
  period. If the upper laser level were in thermal equilibrium with its
  injector level, one would expect  $n_i=n_ue^{-\hbar\omega_{\mathrm
      opt}/k_BT}$, which is a lower bound for the population of the lower
  laser level, $n_l$, being in resonance with the injector level.}
\label{TabLaserParameter}
\end{table}

\begin{figure}
\includegraphics[width=8.5cm]{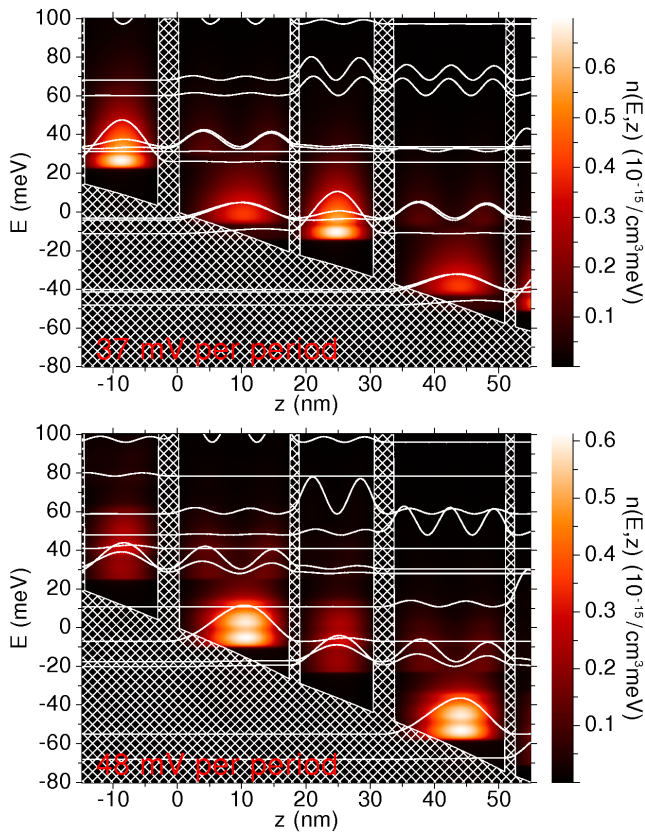}
\caption{(Color online) Electron density at 200 K 
for the first current peak at 37 mV per period (upper panel) and the lasing
operation point at 48 mV per period
 (lower panel). In addition the
 electronic eigenstates are shown, which clearly show the mixing of the Wannier
 states at the respective resonances.}
\label{FigDensity}
\end{figure}

In order to understand the operation, the energetically and spatially resolved
carrier density \cite{LeePRB2006} is shown in Fig.~\ref{FigDensity}.
At 37 mV per period, the extraction level is aligned with the upper laser
level, which causes a pronounced current peak, while a significant part of the
electrons is trapped in the lower laser level. Increasing the bias, the lower
laser level is emptied at 48 mV due to its alignment with the extraction
level. For low temperatures (e.g. 100 K) the scattering 
lifetime of the upper laser level
is long and thus almost all electrons are collected in this level, see the
data given in Table~\ref{TabLaserParameter}. Thus the current is determined by
the tunneling from the upper laser level into the extraction level which
causes NDC at the bias of 48 mV per period. With increasing temperature,
scattering becomes stronger leading to a higher occupation of the lower laser
level, so that the contribution of the tunneling resonance between lower laser 
level and the extraction level is of larger importance. This provides PDC for
each period as observed at 200 and 300 K and required for stable
operation. Thus the scattering from the upper to the lower laser level, which
reduces the inversion (see \cite{JirauschekJAP2007} for a detailed 
discussion), is actually required for the device operation 
in this design. A key feature is the efficient extraction, which maintains the
inversion. Thus the design can be considered as extraction-dominated.

The reduction of gain with temperature  can be attributed to a
combination of increased broadening of the gain profile and reduced 
inversion, see Table~\ref{TabLaserParameter}. The data indicates that thermal
backfilling contributes to the increase of $n_l$ with temperature, but can
only explain half the magnitude. The depopulation kinetics is of
equal relevance. A principle advantage of the new design is that thermal
backfilling can not entirely destroy the inversion as the upper laser
level plays the role of the reservoir, having the highest occupation at all
temperatures. This is a common feature of properly designed scattering
injection lasers.

In conclusion, a new design for THz-QCLs has been proposed based 
on the efficient depopulation of the lower laser level, 
which also ensures the positive differential conductivity.
Lasing operation above 200 K is predicted.

\begin{acknowledgments}
I want to thank S.-C.~Lee and R.~Nelander for their contributions to the
development of the computer code. Discussions with E. Dupont, 
S. Fathololoumi, S. Kumar, and G. Scalari clarified several issues addressed here. Financial support by the Swedish Research Council (VR) is acknowledged.
\end{acknowledgments}


\end{document}